\documentclass{XrU2005}
\usepackage{graphicx}

\title{Principal component analysis of the X-ray variability in NGC 7469}
\author[1]{A. J. Blustin}
\author[1,2]{S. V. Fuerst}
\author[1]{G. Branduardi-Raymont}
\author[1]{M. J. Page}
\author[3]{E. Behar}
\author[4]{J. S. Kaastra}
\affil[1]{UCL Mullard Space Science Laboratory, Holmbury St. Mary, Dorking, Surrey, RH5 6NT, UK}
\affil[2]{KIPAC, Stanford University, Menlo Park, CA 94025, USA}
\affil[3]{Physics Department, Technion, Haifa 32000, Israel}
\affil[4]{SRON National Institute for Space Research, Sorbonnelaan 2, 3584 CA Utrecht, The Netherlands}

\begin{document}


\maketitle

\begin{abstract}
We apply Principal Component Analysis (PCA) to study the variability of the X-ray continuum in the Seyfert 1 galaxy NGC~7469. The PCA technique is used to separate out linear components contributing to variability between multiple datasets; the technique is often used in analysis of optical spectra, but has rarely been applied to AGN X-ray spectroscopy. Running a PCA algorithm on 0.3$-$10~keV EPIC data from a 150~ks XMM-Newton observation of NGC~7469, we describe the spectral components extracted and evaluate the usefulness of the PCA technique for understanding the X-ray continuum in AGN.
\end{abstract}

\section{Introduction and methodology}

Principal Component Analysis (PCA) is a technique which finds linear combinations of the spectral components that produce most of the variability in a time series of spectra. Using only the first few of these components (the `principal ones') one can describe the time series with a greatly reduced number of parameters. These components may correspond to physical parameters of the system that are changing, so the method may be used to discover how many parameters there are, and how they change with time.

In order to calculate the components, first the mean of each variable is subtracted.  This corresponds to the zero$^{\rm th}$ order component.  Next, each variable is scaled by its standard deviation. Generally, the correlation matrix of these shifted, scaled variables is then obtained. The mathematics is simplified, however, by using the method of singular value decomposition to find the eigenvalues and eigenvectors. With this technique, the correlation matrix is not explicitly required, and the raw data matrix is used instead \citep{mittaz1990}. The output of the PCA consists of a set of principal components, a matrix of coefficients and a set of eigenvalues. Each input spectrum can be reconstructed from the components once they have been multiplied by the appropriate set of coefficients which determine how much of an effect each component has in a given spectrum.

A problem with Principal Component Analysis is that it strictly only works when the signal is a superposition of linear components. If there is a non-linear interaction between components then it will `incorrectly' separate them. For example, if the time series consists of the spectrum of a Gaussian line profile which changes width over time, then the components will be similar to Fourier modes. These modes can be combined to produce a line of any width, but do not individually correspond to physical reality. If there is a combination of non-linear effects operating simultaneously within the system, then interpretation can be difficult. Statistical noise in the data is another potential source of confusion. This introduces extra principal components to describe the noise fluctuations, which obviously need to be removed as they do not describe physical processes of interest. This can be done by truncating the list of components in the correct place, by assuming that the weaker components (those with smaller eigenvalues) correspond to the Fourier modes for the broad-band noise.

\begin{figure}
\centering
\includegraphics[width=5.5cm,angle=-90]{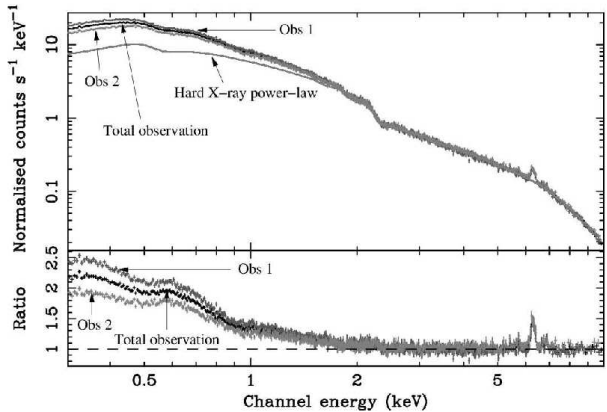}
\caption{EPIC-pn spectrum of NGC~7469 from the first and second part of the observation, and from the whole observation combined. Also the ratios of these spectra to a Galactic-absorbed power-law model fitted to the 3$-$5.5~keV and 7$-$10~keV ranges.}
\label{pn_spec}
\end{figure}

NGC~7469 is a bright nearby Seyfert (z = 0.0164) which was observed for 150~ks by XMM-Newton in November/December 2004. The total exposure time was split into two parts over consecutive orbits. This long observation was obtained primarily for the purposes of high resolution spectroscopy with the RGS; this spectrum contains evidence of absorption by outflowing material with at least two different levels of ionisation. In order to correctly model the soft X-ray spectral features, we need to understand the continuum underlying them. Fig.~\ref{pn_spec} shows the EPIC-pn spectra from the whole of the observation and from both parts separately. The solid line is a power-law model with galactic absorption fitted to the 3$-$5.5~keV and 7$-$10~keV ranges. The lower panel of this plot shows the ratios of the three observed spectra to the power-law fit. The soft excess is clearly variable, so we applied PCA to see if it could separate out different varying components in the EPIC-pn spectrum \citep[see~e.g.][]{vaughan2004}. We ran a PCA code (based on that of \citealt{francis1999}) on a series of twelve $\sim$ 10~ks EPIC-pn spectra (six from each part of the observation). This exposure time was chosen as a compromise between obtaining good time resolution to look for changing components in the spectrum and having sufficient signal-to-noise in each spectrum.

\section{Results and discussion}

\begin{figure}[h]
\centering
\includegraphics[width=6cm,angle=-90]{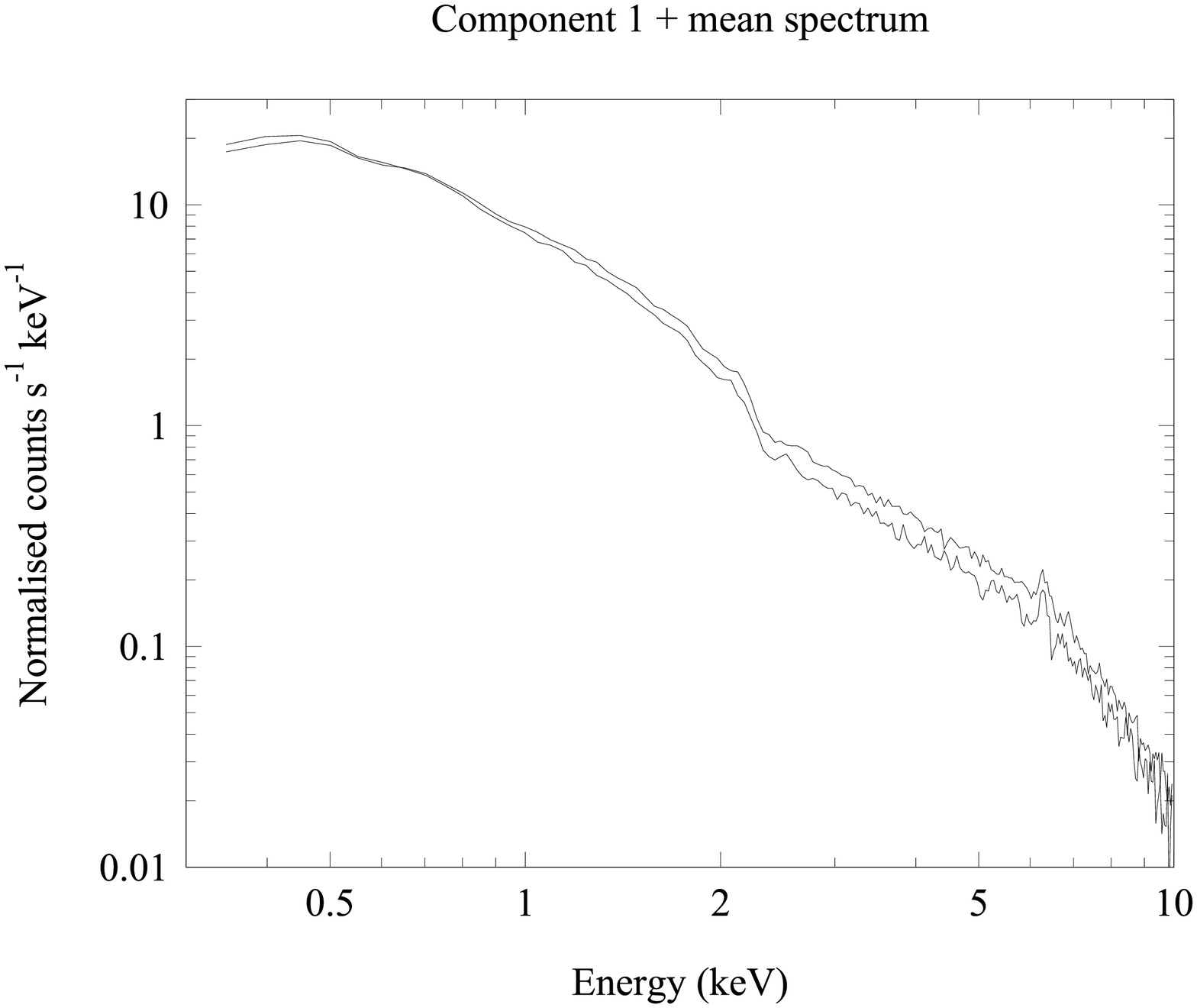}
\caption{Maximum and minimum spectra (added to the mean) for the most variable component.}
\label{pca_comp1}
\end{figure}

\begin{figure}[h]
\centering
\includegraphics[width=6cm,angle=-90]{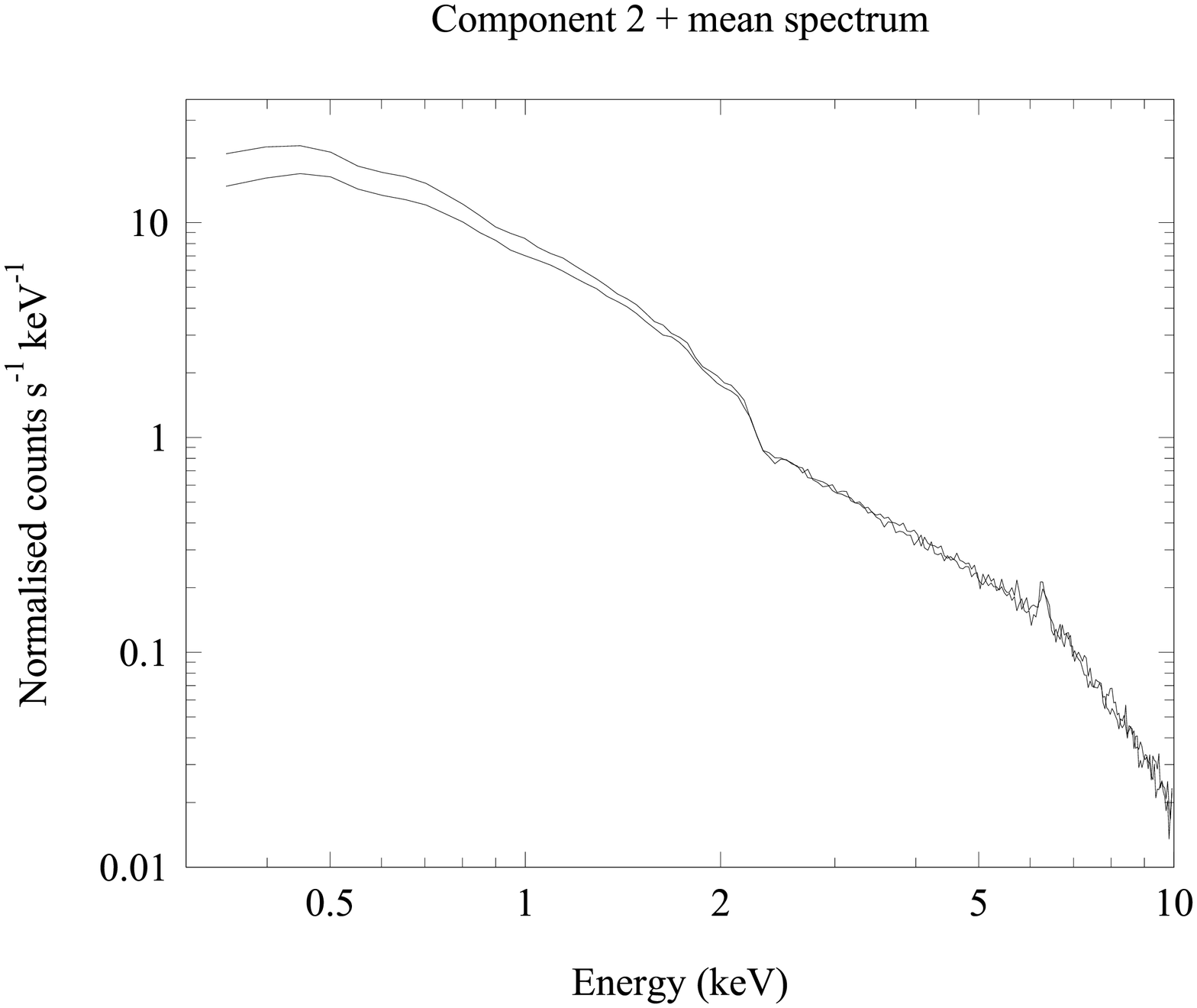}
\caption{Maximum and minimum spectra (added to the mean) for the second most variable component.}
\label{pca_comp2}
\end{figure}

The PCA code generates as many principal components as there are input datasets, in this case twelve. The eigenvalues for each component indicate how much of the spectral variability that particular component is responsible for. In this case, the first two principal components seem to be significant, being responsible for $\sim$18\% and $\sim$13\% of the variability respectively. The other ten components appear to be noise. Fig.~\ref{pca_comp1} and \ref{pca_comp2} show components 1 and 2 added to the mean spectrum (for clarity, only the components plus mean showing the greatest extent of the variability are plotted). These plots indicate the presence of a variable hard power-law and a variable soft power-law, with the hard power-law being more variable than the soft. The iron K$\alpha$ line seems only to be present in the constant mean spectrum but not in the individual components, thus implying that it is not variable on these timescales $-$ or that any variability is below the level of the statistical noise in the input spectra.

Since PCA will reduce a non-linear variability to a series of Fourier-type components, are these two components genuinely physical or simply Fourier deconstructions of a much more complicated reality? A good sign may be that there are only two apparently non-noise components; if the spectrum had been varying in a more complicated way, a larger number of power-law type components might have been generated than the two plausible ones we see here. On the subject of whether PCA tells us anything about the spectrum of NGC~7469 that we did not already know via less sophisticated methods (there is a soft and a hard component, varying in different ways), we conclude that it does not provide us with new information; the method does however give a model-independent confirmation of current knowledge.


\begin{thebibliography}{}
\bibitem[Francis~\&~Wills(1999)]{francis1999} Francis, P. J. \& Wills, B. J. 1999, in Quasars and Cosmology, ASP Conference Series 162, p. 363
\bibitem[Mittaz~et~al.(1990)]{mittaz1990} Mittaz, J. P. D., Penston, M. V. \& Snijders, M. A. J. 1990, MNRAS, 242, 370
\bibitem[Vaughan~\&~Fabian(2004)]{vaughan2004} Vaughan, S. \& Fabian, A. C. 2004, MNRAS, 348, 1415
\end{thebibliography}
\end{document}